\documentclass[reprint,bibnotes,aps]{revtex4-2}
\usepackage{siunitx}
\usepackage{xcolor}
\usepackage{graphicx}
\usepackage{dcolumn}
\usepackage{bm}
\usepackage{xspace}
\usepackage{xpatch}
\usepackage{amsmath}
\usepackage{amssymb}
\usepackage{tikz}

\def\cE     {\ensuremath{\mathcal{E}}\xspace}
\def\mL     {\ensuremath{\mathrm{L}}\xspace}
\def\mA     {\ensuremath{\mathrm{A}}\xspace}
\def\md     {\ensuremath{\mathrm{d}}\xspace}
\def\ag     {\ensuremath{a_n}\xspace}
\def\ae     {\ensuremath{b_n}\xspace}
\def\pg     {\ensuremath{|\ag|^2}\xspace}
\def\pe     {\ensuremath{|\ae|^2}\xspace}
\def\pt     {\ensuremath{P}\xspace}
\def\mA     {\ensuremath{\mathrm{A}}\xspace}
\def\md     {\ensuremath{\mathrm{d}}\xspace}
\def\tK     {\ensuremath{\widetilde{K}}\xspace}
\def\tJ     {\ensuremath{\widetilde{J}}\xspace}
\def\oI     {\ensuremath{\overline{I}}\xspace}
\def\otau   {\ensuremath{\overline{\tau}}\xspace}
\def\ot     {\ensuremath{\overline{t}}\xspace}
\def\oN     {\ensuremath{\overline{N}}\xspace}

\begin{document}
\title{The resonance fluorescence cascade of a laser-excited two-level atom}

\author{Serge Reynaud}
\affiliation{ Laboratoire Kastler Brossel, Sorbonne Universit\'e, CNRS, \\
 ENS-PSL Universit\'e, Coll\`ege de France, Campus Jussieu, F-75252 Paris}

\date{\today}

\begin{abstract}
The cascade of fluorescence photons by a two-level atom excited by coherent laser light is reviewed. The discussion emphasizes the random nature of resonance fluorescence and uses the distribution of delays between two successively emitted photons as the primary characterization of the process. Other characterizations such as photon counting and photon correlation are deduced. 
\end{abstract}

\maketitle

\section{Introduction}
\label{sec:introduction}

Statistical physics of interaction processes between matter and radiation have been at the core of quantum theory since its very beginning \cite{Planck1900,Einstein1905,Einstein1909,Einstein1917,Dirac1927, Weisskopf1930,Fermi1932}. These processes show novel features when an atom is shined by a coherent laser beam tuned in the vicinity of an atomic resonance. Change of frequencies is observed in the fluorescence spectrum \cite{Schuda1974,Wu1975,Grove1977,OrtizGutierrez2019}, while non trivial photon correlation and photon counting show up in the time series of emitted fluorescence photons \cite{Kimble1977,Dagenais1978,Mandel1979,Short1983}. 

As this \emph{resonance fluorescence} problem involves relaxation phenomena, the associated properties have been most often studied by master equation techniques dealing with the density matrix associated with the system \cite{Mollow1969,Carmichael1976,Carmichael1976a,Kimble1976,CohenTannoudji1977h}. Though efficient for computing the fluorescence signals, this method is not well suited to an intuitive understanding of the statistical properties of the emitted light. In particular, the intrinsic randomness of spontaneous emission processes is not emphasized by the master equation techniques \cite{Pomeau2016,Pomeau2021,Pomeau2022}.

Another method for the study of resonance fluorescence is to consider the \emph{dressed atom}, that is the system combining the atom and laser photons. The spectral distribution of fluorescence is understood immediately by looking at the eigenstates of this dressed atom \cite{CohenTannoudji1977}. The statistical properties of the fluorescence are interpreted from the radiative cascade of the dressed atom going downwards its energy diagram while spontaneously emitting the fluorescence photons \cite{CohenTannoudji1979,Reynaud1981,Reynaud1983}.

The present review is devoted to the study of the cascade of the fluorescence photons emitted by a two-level atom excited by a nearly resonant laser. This study gives a comprehension of its statistical properties fully accounting for the intrinsic randomness of spontaneous emission. The primary characterization is the delay function describing the distribution of intervals between successively emitted fluorescence photons. Explicit expressions of this function are given for arbitrary values of the parameters and other characterizations of the fluorescence statistics are derived from it. 
The results of the method presented here are compared with those found in the existing literature, which have in some cases been tested against the results of dedicated experiments. 

\section{Resonance fluorescence cascade}
\label{sec:cascade}

We first introduce the states of the dressed atom, labeled by two quantum numbers, an atomic one ($g$ for the ground state and $e$ for the excited one) and the number $n$ of laser photons. The states $\left| g , n \right>$ and $\left| e , n-1 \right>$ form a nearly degenerate twofold multiplicity $\cE_n$. The splitting of these two states, measured as an angular frequency, is the detuning $\delta = \omega_\mL - \omega_\mA$ of the laser frequency $\omega_\mL$ with the atomic one $\omega_\mA$. 

The two states in $\cE_n$ are coupled by the interaction describing absorption and stimulated emission of laser photons by the atom, with the coupling, the Rabi frequency $\Omega$, proportional to the laser field amplitude. The parameters $\delta, \Omega$ are much smaller than the separation $\omega_\mL$ between adjacent multiplicities, so that the couplings between different multiplicities can be safely disregarded. The variation with $n$ of $\Omega$ (proportional to $\sqrt{n}$) is also neglected, as the number of fluorescence photons remains much smaller than the huge number of laser photons.

Spontaneous emission processes, due to the coupling of the atom with vacuum fluctuations in all field modes, is described by transitions downwards the diagram of dressed atom states. 
The parameter characterizing the spontaneous processes is the Einstein coefficient $\Gamma$ appearing in front of populations in rate equations \cite{Einstein1917} or the decay parameter for atomic coherences  $\gamma=\Gamma/2$ fixing the width of fluorescence lines \cite{CohenTannoudji1977h}.

\begin{figure}[t!]
\includegraphics[scale=0.6]{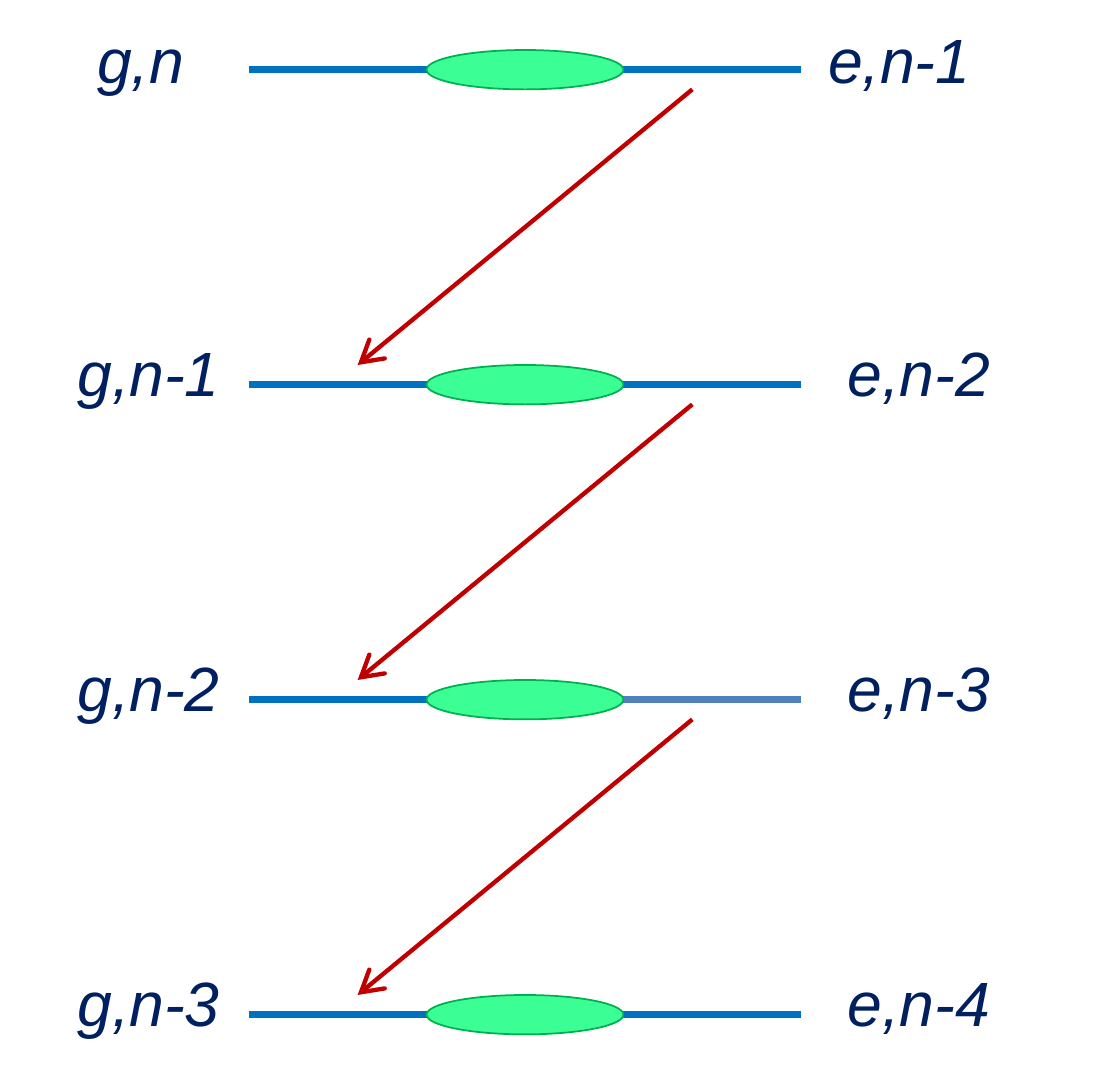}
\caption{Energy diagram of the atom ($g$ for the ground state, $e$ for the excited one) dressed by laser photons (number $n$). States (blue lines) are grouped in twofold nearly degenerate multiplicities in which they are coupled by reversible absorption and stimulated emission processes (green shaded ellipses). Spontaneous emission processes are irreversible jumps downward the ladder of multiplicities (red arrows). }
\label{fig:cascade}  
\end{figure}

The fluorescence process appears as a succession of elementary processes sketched on the diagram of Fig.\ref{fig:cascade}. Absorption and stimulated emission of laser photons by the atom are described by the reversible evolution in a given multiplicity $\cE_n$ while spontaneous emission processes correspond to irreversible jumps from a multiplicity $\cE_n$ to the adjacent lower one $\cE_{n-1}$. The random jumps interrupt the reversible evolution while projecting the atom from the excited to the ground state \cite{CohenTannoudji1979}.

With the dependence of $\Omega$ versus $n$ neglected, the diagram of Fig.\ref{fig:cascade} is stationary under changes of $n$, and it is tempting to fold up the multiplicities on a density matrix describing the atom-field system, thus reproducing the master equation commonly used to study resonance fluorescence \cite{Mollow1969,Carmichael1976,Carmichael1976a,Kimble1976,CohenTannoudji1977h}. The drawback of this method is that the reversible and irreversible processes are mixed up in the folding process, resulting in a dilution of their intrinsic difference. 

A much better way to proceed is to go on with the full diagram of Fig.\ref{fig:cascade} which keeps the trace of the number of fluorescence photons. The full process is called a radiative cascade because of its analogy to cascades happening in atomic or nuclear configurations with successive transitions \cite{Popp1970}.
It has been described by master equation techniques derived from general relaxation theory \cite{Reynaud1981,Reynaud1983}, also using the sophisticated methods of photodetection theory \cite{Glauber1963,Kelley1964,Davidson1968}.
Here we use a more intuitive approach which gives direct access to the distribution of delays between successively emitted photons, while producing the same results as master equation techniques. 
 
The evolution is written for a wave function defined in a multiplicity $\cE_n$ (the initial value of $n$ being irrelevant thanks to the stationarity of the diagram of Fig.\ref{fig:cascade})
\begin{equation}
\begin{aligned}
&\left| \Psi_n \left(\tau\right) \right> = 
\ag \left(\tau\right) \left| g , n \right> 
+ \ae \left(\tau\right) \left| e , n-1 \right>   ~,\\
&\ag^\prime = -\imath \frac\Omega2 \ae  ~, \quad
\ae^\prime = -\imath \frac\Omega2 \ag 
-  \left(\gamma-\imath\delta  \right) \ae ~,
\end{aligned}
\label{eq:evolution}
\end{equation}
where the prime symbol denotes a derivation with respect to time $\tau$.
With the initial state set just after the detection of a fluorescence photon, that is also just after a projection in the ground state
($\ag(0)=1$, $\ae(0)=0$), these equations are solved as 
\begin{equation}
\begin{aligned}
\ag &= \left( \cosh \frac{\chi \tau}{2} 
+ \frac{\gamma -\imath \delta }{\chi} \sinh \frac{\chi \tau}{2}\right) 
e^ {-\left(\gamma -\imath \delta \right) \frac{\tau}{2} }~, \\
\ae &= - \frac{\imath \Omega}{\chi}  \sinh \frac{\chi \tau}{2} 
e^ {-\left(\gamma -\imath \delta \right) \frac{\tau}{2} }~, \\
\chi &  = \sqrt{\left(\gamma -\imath \delta \right)^2-\Omega^2} ~.
\end{aligned}
\label{eq:solution}
\end{equation}

We also define the populations in the two states 
of the multiplicity $\cE_n$
\begin{equation}
\begin{aligned}
 \pg  &= 
\left| \cosh \frac{\chi \tau}{2}
+ \frac{\gamma -\imath \delta }{\chi} 
\sinh \frac{\chi \tau}{2} \right|^2 e^{-\gamma \tau}~, \\
 \pe  &=
\left| \frac{\Omega}{\chi} 
 \sinh \frac{\chi \tau}{2} \right|^2 e^{-\gamma \tau}~.
\end{aligned}
\label{eq:population}
\end{equation}
At resonance, $\chi$ is real for $\Omega<\gamma$ and purely imaginary for $\Omega>\gamma$. 
The forms (\ref{eq:solution}-\ref{eq:population}) have to be taken as the limits for $\chi\to0$ at the border $\Omega=\gamma$ (for $\delta=0$).

We now define the delay function $K$ (called the waiting function $W$ in \cite{Reynaud1988}) which is the probability of emission of the next fluorescence photon at time $\tau$ after an emission at time 0. It is the product of the Einstein coefficient $\Gamma=2\gamma$ by the population $\pe$ calculated in \eqref{eq:population} for an atom in the ground state at $\tau=0$
\begin{equation}
K \left(\tau\right) = 2\gamma \pe  
= - \pt ^\prime \left(\tau\right)\,,  \quad
\pt\left(\tau\right)= \pg  + \pe ~.
\label{eq:defK}
\end{equation}
The fact that $K$ is the opposite of time derivative of the total population $\pt$ in $\cE_{n}$ is an immediate consequence of equations \eqref{eq:evolution}. It means that $K$ is the feeding rate of population in the adjacent lower multiplicity $\cE_{n-1}$, starting a new reversible evolution in $\cE_{n-1}$ which leads to the next fluorescence event.
Equation \eqref{eq:defK} implies that the cumulative distribution function for the emission of the next fluorescence photon is $1-\pt(\tau)$, going from 0 to 1 when $\tau$ runs from 0 to $\infty$. Hence, the intensity $K(\tau)$ has a unity integral when integrated over $\tau$. 

We stress at this point that the fluorescence process is intrinsically random. As we consider only signals built on the fluorescence intensity,  
the process can be understood as a series of random times of emission of successive photons. 
Each of the independent random delays are characterized by the delay function $K$ or, equivalently, by a non stationary Poisson point process \cite{Pomeau2016,Pomeau2021,Pomeau2022}
\begin{equation}
\begin{aligned}
&\Lambda \left(\tau\right) =\ln\frac1{\pt\left(\tau\right)} ~,  \quad
\lambda \left(\tau\right) \equiv \Lambda^\prime \left(\tau\right) 
=\frac{K\left(\tau\right)}{\pt\left(\tau\right)} ~,\\
&K \left(\tau\right) = \lambda \left(\tau\right)
\exp\left(-\Lambda \left(\tau\right)\right) ~.
\end{aligned}
\label{eq:defLambda}
\end{equation}
The non stationary Poisson intensity $\lambda(\tau)$ is defined from $K(\tau)$ after a division by the total population $\pt(\tau)$ still in $\cE_n$ at time $\tau$. It would be constant for a normal Poisson process, but this is not the case here as $\lambda(\tau)$ starts from a null value at $\tau=0$.

\section{Photon counting statistics}
\label{sec:counting}

We now discuss the photon counting statistics which has been studied theoretically \cite{Mandel1979,Reynaud1983,Kim1987,Carmichael1989,Kim1989,Davidovich1996} and experimentally \cite{Short1983,Schubert1992,Fleury2000,Treussart2004}. It reveals non trivial statistics of the number $N_T$ of fluorescence photons emitted in a given time interval $T$, which may possibly be sub-Poissonian, that is more regular than standard Poisson statistics. 

Here, we derive the counting statistics from the delay function \cite{Reynaud1981,Reynaud1983,Reynaud1988}, using the relations discussed for more general point processes in \cite{Reynaud1990}. We first study the random variable $t_n$ defined as the time of emission of the $n-$th photon after the emission of a $0-$th photon. With $t_0$ set to 0, $t_n$ is the sum of $n$ successive delays (as far as possible, we use $\tau$ for delays between successive emissions and $t$ for emission times) 
\begin{equation}
t_n \equiv \tau_1 + \tau_2 + \ldots + \tau_n ~.
\label{eq:defTn}
\end{equation}

\begin{figure}[t!]
\includegraphics[scale=0.5]{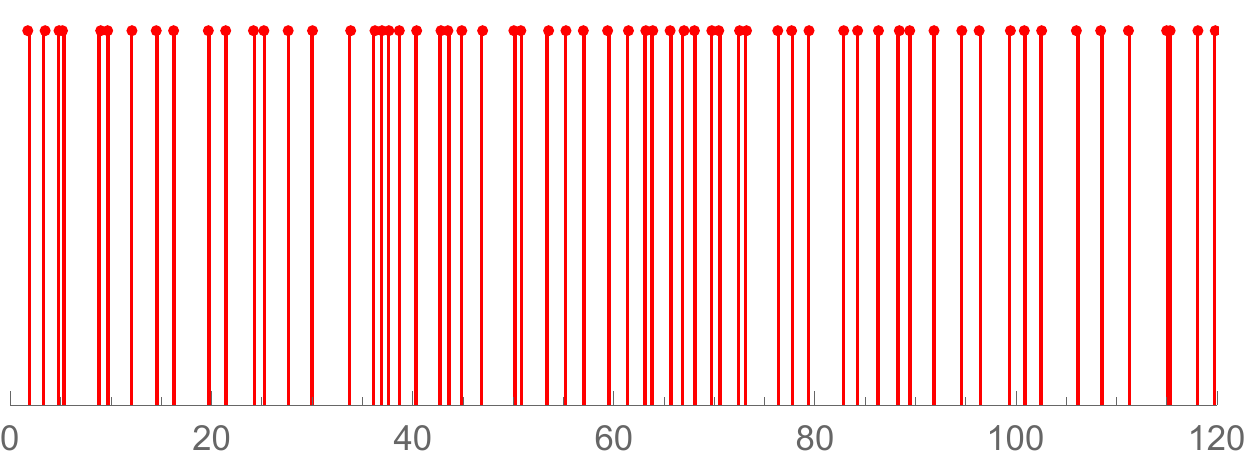}
\includegraphics[scale=0.5]{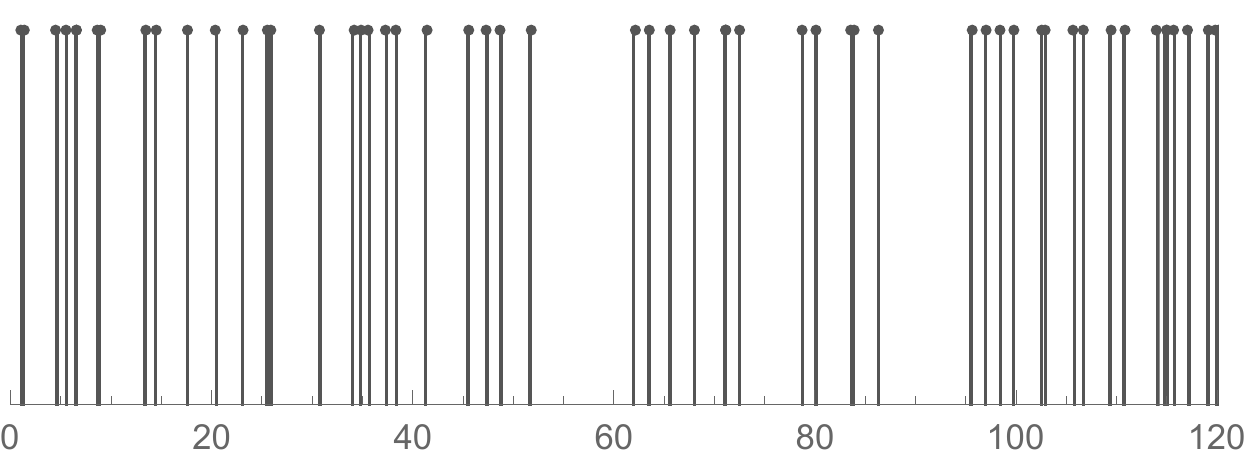}
\caption{Random draw of a series of fluorescence times (red, upper plot) for $\Omega=\sqrt{2}\gamma,\delta=0$, compared to a standard Poisson process with the same mean delay (gray, lower plot). Time is measured by the dimensionless variable $\gamma t_n$. }
\label{fig:random}  
\end{figure}

\begin{figure}[t!]
\includegraphics[scale=0.6]{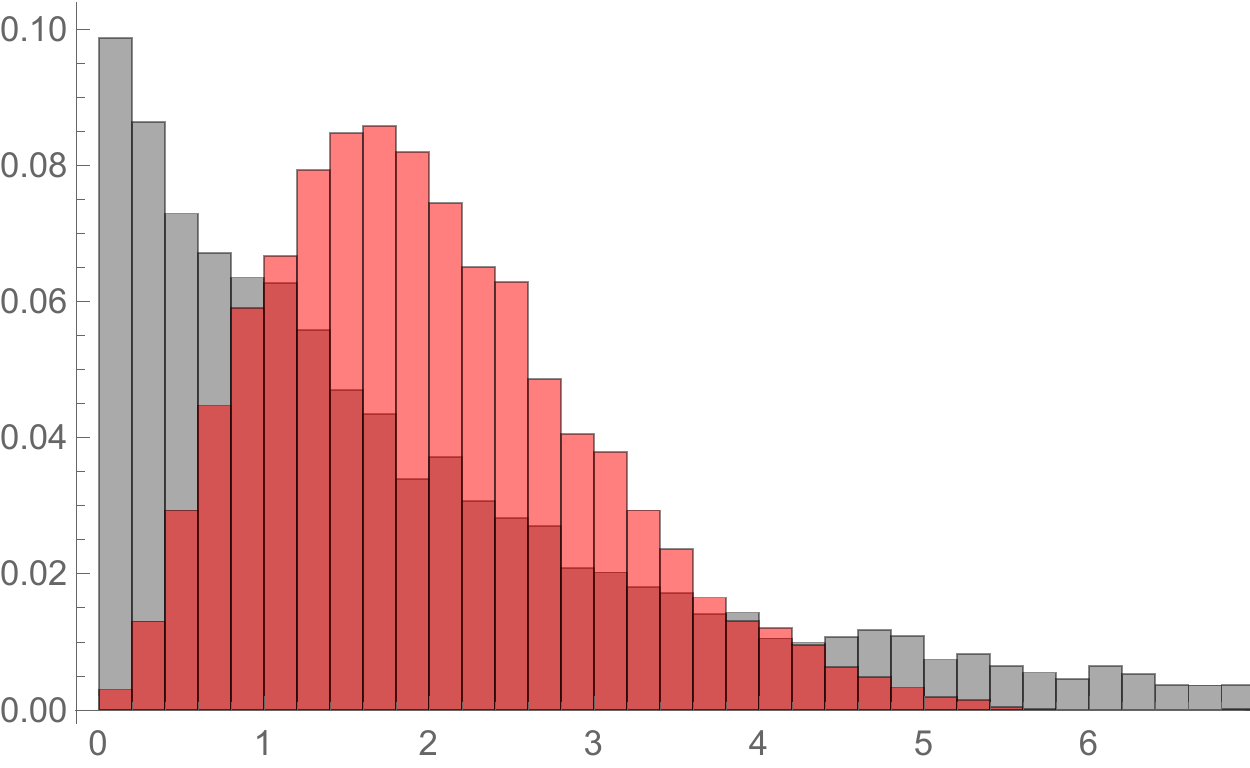}
\caption{Probability per channel for 10000 random delays drawn for $\Omega=\sqrt{2}\gamma,\delta=0$ (red), compared to a standard Poisson process with the same mean delay (gray). Delay is measured by the dimensionless variable $\gamma \tau$. }
\label{fig:simkfunction}
\end{figure}

An example of series of $t_n$ is shown on the upper plot of Fig.\ref{fig:random} by successive red bars, with random delays drawn with the distribution $K$ calculated for $\Omega=\sqrt{2}\gamma$ and $\delta=0$.
Each delay is obtained as $\pt^{(-1)}(u)$ where $\pt^{(-1)}$ is the inverse function of $\pt$ while $u$ is a random variable with a uniform distribution on $\left[0,1\right]$. The parameters are chosen so that the flow of photons shows a good regularity, with a large noise reduction compared to the standard Poisson noise. On Fig.\ref{fig:random}, the regularity is seen through a comparison with a Poissonian flow on the lower plot (calculated with the same mean delay). A clearer representation of the regularity property is shown by the histograms on Fig.\ref{fig:simkfunction}, corresponding to random draws of 10000 delays in the sub-Poissonian and Poissonian distributions. 
Other characterizations will be discussed below.

As the different delays are independent random variables with identical probability distribution $K$, the distribution $K_n$ of $t_n$ is given by a repeated convolution on $K$ ($\otimes$ representing a convolution product)
\begin{equation}
K_1 \equiv K~, \quad K_2 \equiv K \otimes K_{1}~, \quad K_n = K \otimes K_{n-1} ~.
\label{eq:defKn}
\end{equation}
This leads to simple algebraic relations between the Laplace transforms of the functions 
\begin{equation}
\begin{aligned}
&\tK \left(s\right) 
= \int_0^\infty e^{-s \tau} K \left(\tau\right) \md \tau~, \\
&\tK_n \left(s\right) = \left(\tK\left(s\right)\right)^n~.
\end{aligned}
\label{eq:laplacetra}
\end{equation}
In particular, one easily gets the mean value and variance of $t_n$ from those of $\tau$ 
\begin{equation}
\ot_n=n\otau ~, \quad \Delta t_n^2= n \Delta \tau^2 ~.
\label{eq:statTn}
\end{equation}

The Laplace transform $\tK(s)$ can be written as a rational function since $K(\tau)$ is a sum of exponential functions 
\begin{equation}
\tK \left(s\right) = \frac{\gamma \Omega ^2 (s+\gamma)}
{s (s+2\gamma) \left((s+\gamma)^2+\delta ^2\right) 
+\Omega ^2 (s+\gamma)^2} ~,
\label{eq:laplaceK}
\end{equation}
with $\tK(0)=1$ as a consequence of the normalization of $K(\tau)$.
The mean $\otau$ and variance $\Delta \tau^2$ of $\tau$ can be computed through derivations of $\ln\tK(s)$ 
\begin{equation}
\begin{aligned}
\otau &= -\left(\ln\tK\right)^\prime_{s=0} 
= \frac{2 \left(\gamma^2+\delta ^2\right) +\Omega ^2}{\gamma  \Omega ^2} 
~, \\
\Delta \tau^2  &= \overline{\tau^2} - \otau^2
=\left(\ln\tK\right)^{\prime\prime}_{s=0} \\
&= \frac{4 \left(\gamma^2+\delta ^2\right)^2 
+2\left(3\delta ^2- \gamma ^2\right) \Omega ^2 +\Omega^4 }
{\gamma ^2 \Omega ^4} ~. 
\end{aligned}
\end{equation}

 We study now the counting statistics of the number $N_T$ of fluorescence photons emitted during a given time interval $T$, assuming that $T$ is long enough so that the number $N_T$ is large. The statistical distribution of $N_T$ thus tends to be Gaussian as a consequence of the central limit theorem. It is then easy to infer the nearly Gaussian counting statistics of $N_T$ from that of $t_n$ with the following mean value and variance \cite{Reynaud1983,Kim1987,Reynaud1988} 
\begin{equation}
\begin{aligned}
&\oN_T=\frac T \otau \,, \quad 
\Delta N_T^2=  \oN_T \frac{\Delta \tau^2}{\otau^2} 
=\oN_T \left(1+Q\right) ~, \\
&Q\equiv\frac{\Delta \tau^2}{\otau^2} -1
=\frac{2\left(\delta ^2-3 \gamma ^2\right) \Omega ^2 }
{\left(2 \left(\gamma ^2+\delta ^2\right)+\Omega ^2\right)^2} ~.
\end{aligned}
\label{eq:mandelQ}
\end{equation}
The variance $\Delta N_T^2$ scales versus $T$ as the mean $\oN_T$, and the ratio between these quantities is usually written in terms of the $Q-$factor introduced by Mandel \cite{Mandel1979}. This factor is interpreted here from its definition in \eqref{eq:mandelQ}. 

Negative values of $Q$ correspond to sub-Poissonian statistics with photon number fluctuations smaller than in standard Poisson statistics, They are obtained when the detuning is not too large ($\delta ^2<3 \gamma ^2$). In particular $Q_0$, $Q$ at resonance ($\delta=0$), is negative for any $\Omega$ 
\begin{equation}
Q_0 =  - \frac{3 \Omega_\mathrm{opt}^2 \Omega ^2 }
{\left(\Omega_\mathrm{opt}^2+\Omega ^2\right)^2}~, \quad
\Omega_\mathrm{opt}=\sqrt{2}\gamma~.
\label{eq:mandelQresonant}
\end{equation}
A value $Q_\mathrm{opt}=-\tfrac34$ is obtained at $\Omega=\Omega_\mathrm{opt}$ which gives in principle an optimal reduction $1+Q_\mathrm{opt}=\tfrac14$ with respect to standard Poisson noise. This corresponds to the regular flow shown on the upper plot of Fig.\ref{fig:random} and on Fig.\ref{fig:simkfunction}.
Experimentally obtained reductions are smaller than that as the detrimental effect of imperfect detection efficiency has to be accounted for \cite{Mandel1979,Reynaud1983}.

\begin{figure}[t!]
\includegraphics[scale=0.5]{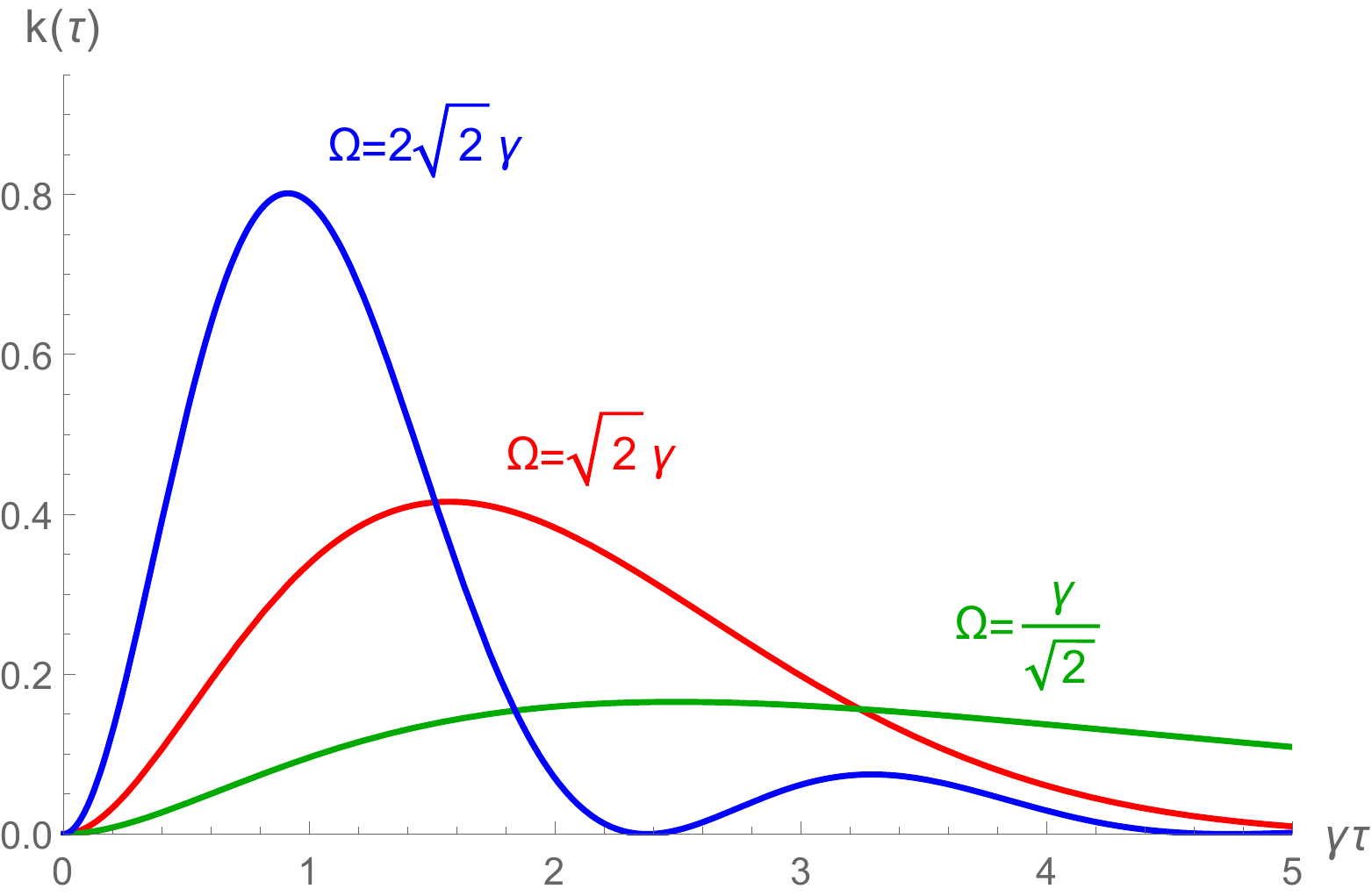}
\caption{Dimensionless function $k=K/\gamma$ drawn versus dimensionless delay $\gamma\tau$ for $\Omega=\Omega_\mathrm{opt}=\sqrt{2}\gamma$ (red), $\Omega=\Omega_\mathrm{opt}/2$ (green) and $\Omega=2\Omega_\mathrm{opt}$ (blue) at resonance ($\delta=0$). }
\label{fig:kfunction}
\end{figure}

We note that the same value $Q_0$ is obtained in \eqref{eq:mandelQresonant} for values of $\Omega$ forming a geometric progression around the optimum $\Omega_\mathrm{opt}$ (for example $1+Q_0=\frac{13}{25}$ for $\Omega=\Omega_\mathrm{opt}/2$ and $\Omega=2\Omega_\mathrm{opt}$). But the functions $K(\tau)$, drawn on Fig.\ref{fig:kfunction} with the optimal case, have different shapes in these two cases, which means that $Q$ is not a full characterization of the photon statistics. A more complete characterization will be given in the next section.

\section{Photon correlation}
\label{sec:correlation}

We then discuss the photon correlation signal which has been studied theoretically \cite{Carmichael1976,Carmichael1976a,Kimble1976,CohenTannoudji1977h} and experimentally \cite{Kimble1977,Dagenais1978}. This correlation reveals the antibunching effect which has been used as a characterization tool for a large variety of systems of physical interest
\cite{Diedrich1987,Basche1992,Schadwinkel2000,Michler2000,Brouri2000,Leibfried2003,Hennrich2005,Lounis2005,Darquie2005,Beugnon2006,Zinoni2006,Press2007,Hagele2008,Gerber2009,Heine2009,Lupton2021,Avriller2021,Darsheshdar2021}.

The intensity correlation signal $C_I (t)$  is associated to the detection of a fluorescence photon at time $t_0$ and another one at time $t_0+t$. It does not depend on $t_0$ because of the stationarity of the mean intensity, so that the discussion is focused on the role of $t$. With $t_0$ set to 0, the time $t$ can be any of the times $t_n$ with $n=1,2,\ldots$, not necessarily the next one as in the definition of $K$. We then define the function $J(t)$ by using the results of the previous section 
\begin{equation}
J\left(t\right) = K_1\left(t\right) + K_2\left(t\right) +  \ldots + K_n\left(t\right) +  \ldots  ~.
\label{eq:defJ}
\end{equation}
This relation takes a simple algebraic form when written in terms of Laplace transforms 
\begin{equation}
\tJ \left(s\right) =\tK + \tK^2 +  \ldots+ \tK^n +  \ldots = \frac{\tK\left(s\right)}{1-\tK\left(s\right)}   ~.
\label{eq:deftJ}
\end{equation}

As $\tK$ is a rational function of $s$, it is also true for $\tJ$
\begin{equation}
\tJ = \frac{\gamma}s \;
\frac{\Omega ^2\left(s+\gamma\right) }
{(s+2\gamma) \left((s+\gamma)^2+\delta ^2\right) 
+\Omega ^2 (s+\gamma)}~.
\label{eq:laplaceJ}
\end{equation}
The function $\tJ(s)$ can itself be decomposed as a sum of elementary rational functions
\begin{equation}
\tJ\left(s\right) = \frac{\oI}s + \Delta \tJ\left(s\right)  ~, \quad
\Delta \tJ\left(s\right) = \sum_{m=1}^3\frac{\rho_m}{s+r_m}~,
\label{eq:decomposeJ}
\end{equation}
where $\oI$ is the mean intensity calculated as the residue associated with the pole of $\tJ$  at $s=0$, that is also the inverse of the mean delay given in \eqref{eq:mandelQ}
\begin{equation}
\oI = \lim_{s\to0} \left(s \tJ \left(s\right) \right) 
= \frac{\gamma  \Omega ^2}{2 \left(\gamma^2+\delta ^2\right) +\Omega ^2}
= \frac1{\otau} ~,
\label{eq:meanintensityl}
\end{equation}
while the $r_m$'s are the opposite of non null roots of the denominator of $\tJ$ and the $\rho_m$'s the associated residues. 

\begin{figure}[t!]
\includegraphics[scale=0.5]{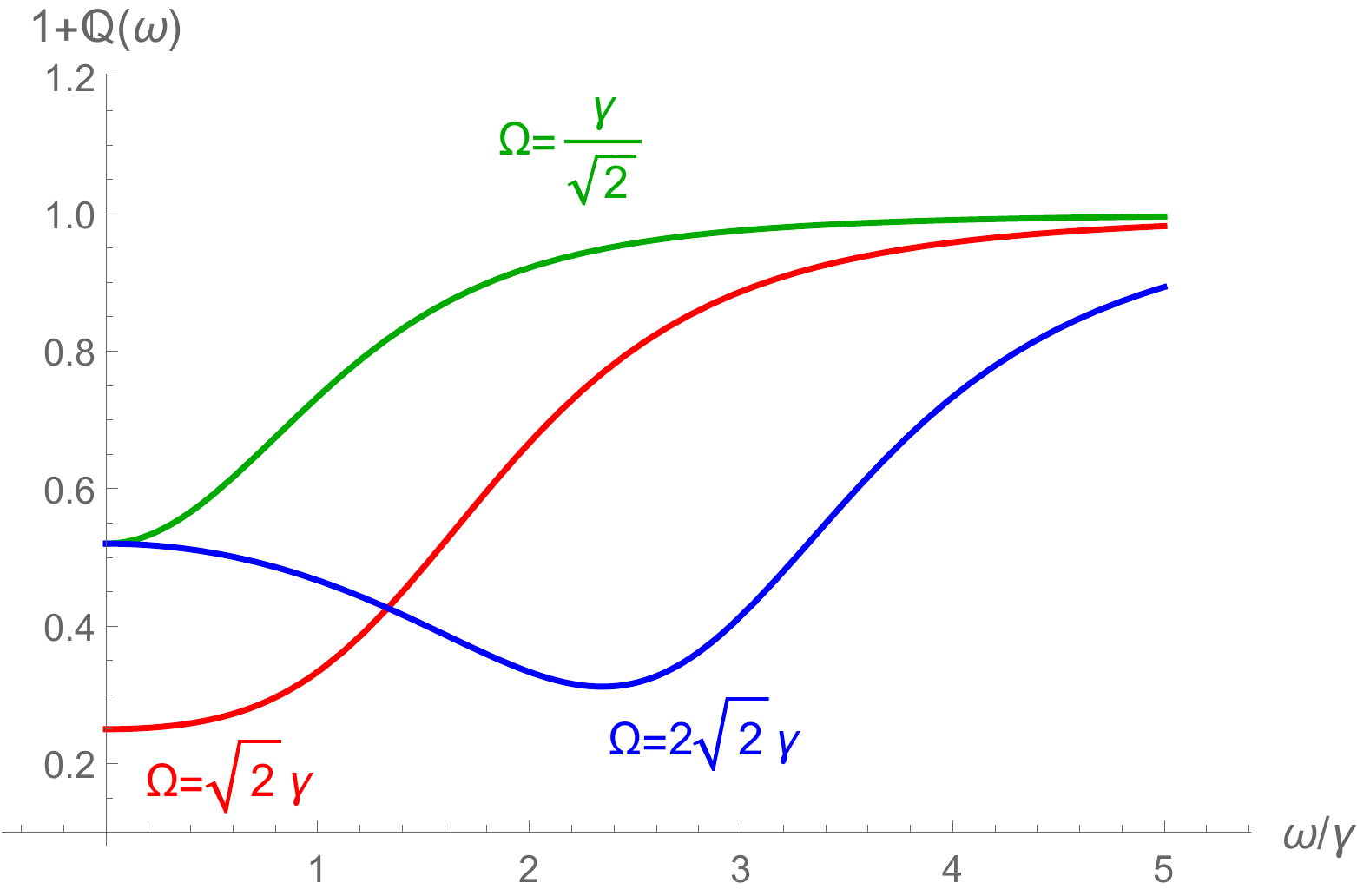}
\caption{Function $1+Q(\omega)$ representing the reduction of standard Poisson noise versus dimensionless frequency $\omega/\gamma$ with the same parameters and color codes as on Fig.\ref{fig:kfunction}.}
\label{fig:qfunction}
\end{figure}

We do not write explicitly the roots and residues which are readily obtained by solving a cubic equation with formal or numerical software. The function $J(t)$ is deduced as a sum (Heaviside functions $\theta(t)$ inserted to emphasize that $J(t)$ and $\Delta J(t)$ are defined for $t\geq0$)
\begin{equation}
\begin{aligned}
&J\left(t\right) = \oI \, \theta\left(t\right) + \Delta J\left(t\right) ~,\\
&\Delta J\left(t\right) = \theta\left(t\right) 
\sum_{m=1}^3 \rho_m e^{-r_m t}~.
\end{aligned}
\label{eq:decomposeJtau}
\end{equation}
The photon correlation function $C_I (t)$ is then obtained in terms of already introduced functions (details in the Appendix B of \cite{Reynaud1990})
\begin{equation}
C_I \left(t\right) = \oI \left( \oI \, \delta\left(t\right) 
+ \Delta J\left(t \right) + \Delta J\left(-t \right)\right)  ~.
\label{eq:correlationI}
\end{equation}
The first term in the parenthesis corresponds to the detection of the same photon at $t_0=0$  and $t$ and it involves a Dirac function $\delta(t)$ describing this simultaneity. The second and third terms represent the detection of photons at $t\neq t_0$ with $t>t_0$ and $t<t_0$ respectively.

Another interesting characterization is the photon noise spectrum $S_I (\omega)$ defined as the Fourier transform of $C_I(t)$ which can be measured by sending the intensity signal to a spectrum analyzer. We obtain for this signal (see the Appendix B of \cite{Reynaud1990})
\begin{equation}
\begin{aligned}
&S_I \left(\omega\right) 
= \oI \left(1 + Q \left(\omega \right) \right)  ~, \\
&Q \left(\omega\right) = \Delta\tJ\left(\imath\omega \right)
+ \Delta\tJ\left(-\imath\omega \right)  ~.
\end{aligned}
\label{eq:spectrumI}
\end{equation}
The function $\Delta\tJ(s)$, written from $\tJ$ and $\oI$
\begin{equation}
\Delta\tJ \left(s\right) = 
 \frac{\oI\left(\gamma ^2+\delta ^2-(s+2\gamma)^2\right)}
 {(s+2\gamma) \left((s+\gamma)^2+\delta ^2\right)+\Omega ^2 (s+\gamma)} ~,
\label{eq:deltaJ}
\end{equation}
is regular at $s=0$ where its value is directly related to the Mandel $Q-$factor, as $Q(0) = 2\Delta\tJ(0)\equiv Q$. 

Hence, $Q(\omega)$ is a generalization of the $Q-$factor, showing sub-Poissonian noise at non zero frequencies. 
Examples of the function $1+Q (\omega)$ are drawn on Fig.\eqref{fig:qfunction} for the same parameters and color codes as on Fig.\eqref{fig:kfunction}. The red curve corresponds to the optimal reduction with $1+Q=\tfrac14$ at null frequency then increasing to reach the standard value 1 at large frequencies. The two other curves correspond to the same value of $Q$ at null frequency, but they show different behaviours as functions of $\omega$, which means that $Q (\omega)$ contains more information on photon statistics than $Q (0)$. 

\begin{figure}[t!]
\includegraphics[scale=0.5]{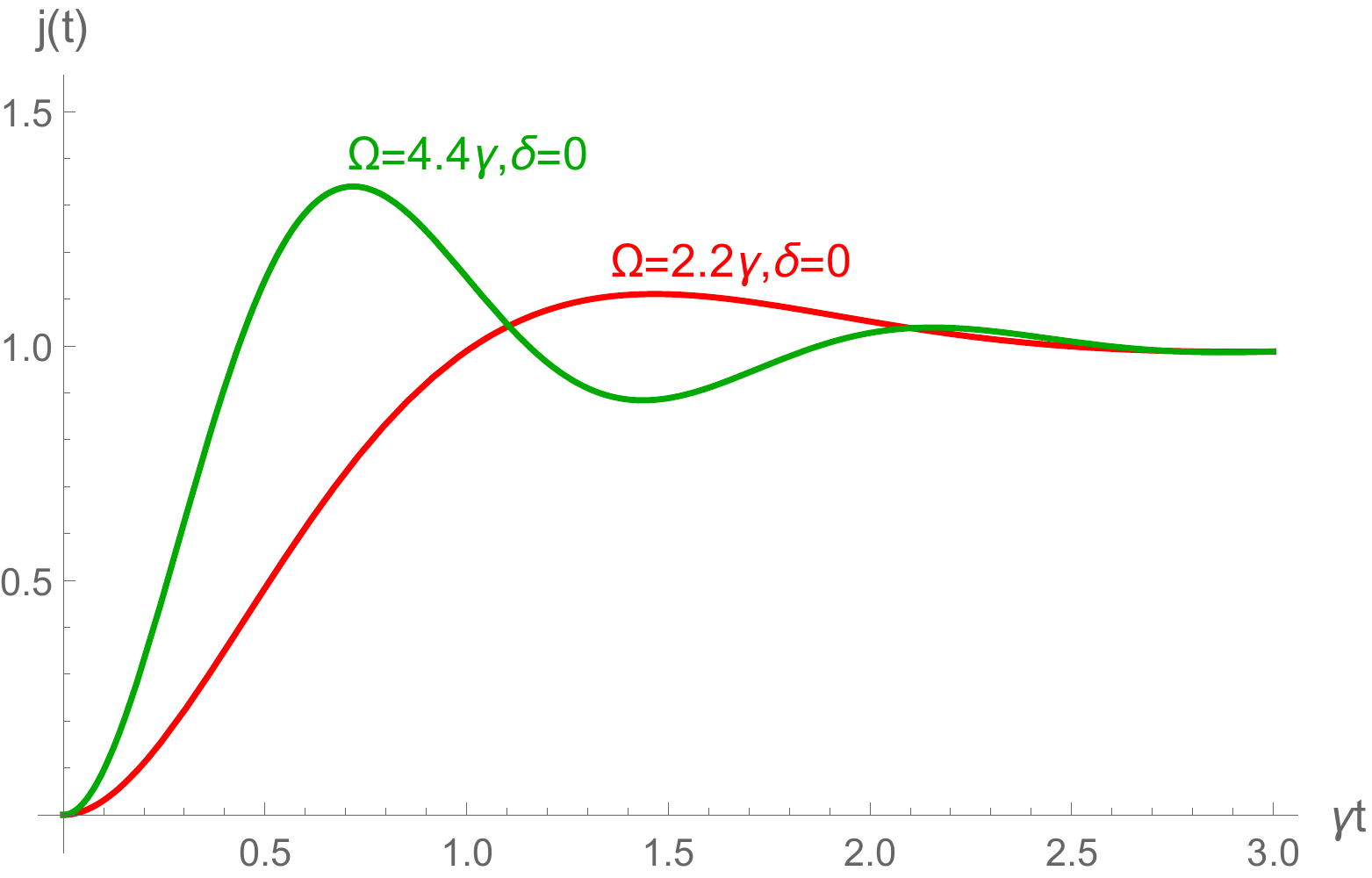}
\caption{Dimensionless function $j=J/\oI$ drawn versus dimensionless time $\gamma t$ for $\Omega=2.2\gamma$ (red) and $\Omega=4.4\gamma$ (green) at resonance ($\delta=0$). These functions match those drawn on Fig.4 of ref.\cite{Dagenais1978} calculated for the same parameters. }
\label{fig:DagenaisFig4}
\end{figure}

Note that sub-Poissonian statistics opened the way \cite{Mandel1982} to the study of squeezing, with a large number of applications, for which we quote a few reviews 
\cite{Walls1983,Loudon1987,Kimble1987,Walls1990,Reynaud1992,Dodonov2002,Schnabel2010,Andersen2016,Schnabel2017,Fabre2020}.

\section{Comparison with literature}
\label{sec:particular}

We now emphasize that the expressions given in this review match those written in many papers in the existing literature, with some of them shown to agree with the results of dedicated experiments.

\begin{figure}[t!]
\includegraphics[scale=0.5]{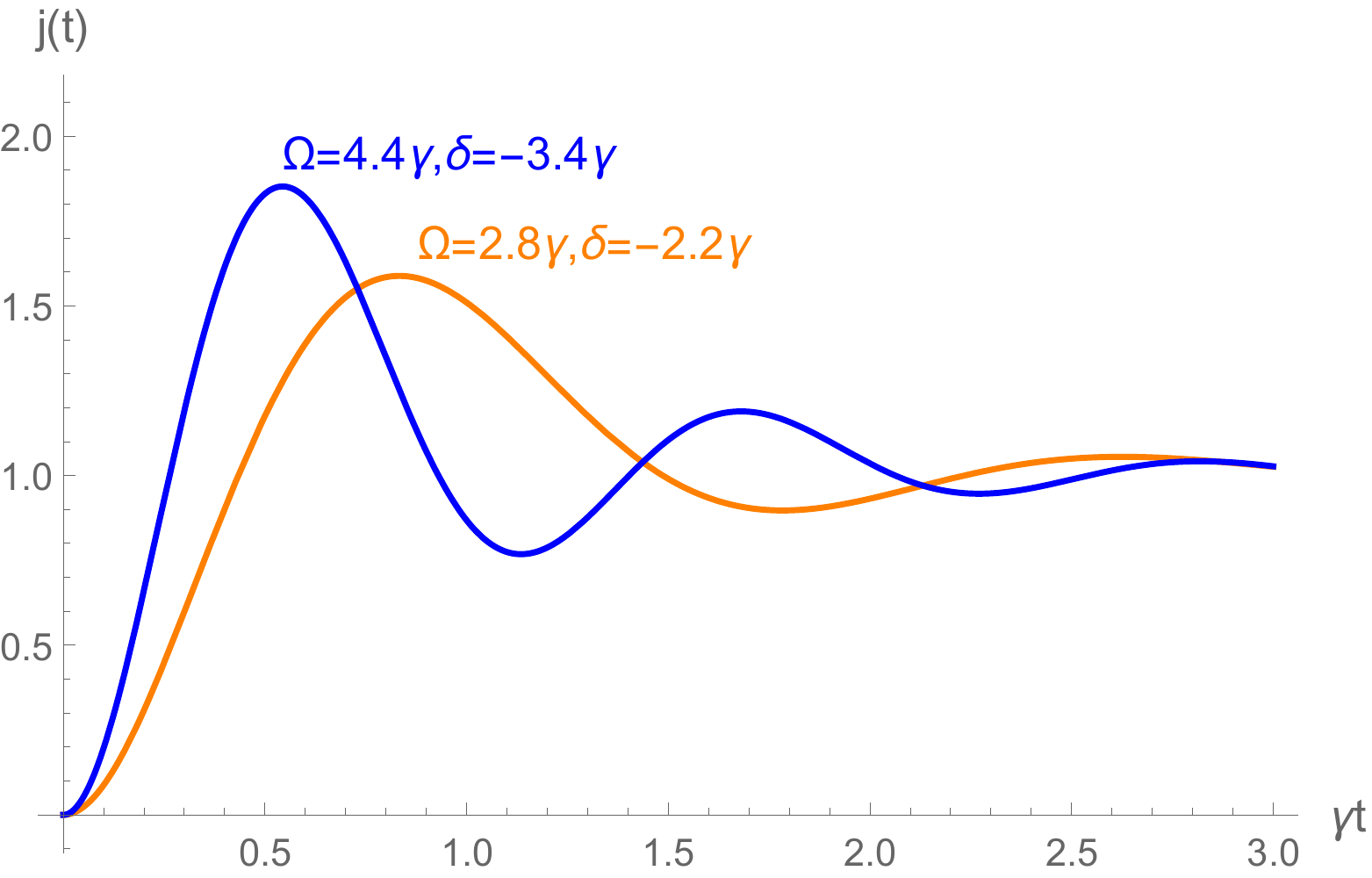}
\caption{Dimensionless function $j=J/\oI$ drawn versus $\gamma t$ for the parameters $\Omega=2.8\gamma,\delta=-2.2\gamma$ (orange) and $\Omega=4.4\gamma,\delta=-3.4\gamma$ (blue). These functions match those on Fig.5 of ref.\cite{Dagenais1978} calculated for the same parameters. }
\label{fig:DagenaisFig5}
\end{figure}

In the case of resonant excitation, $j_0$ ($j\equiv J/{\oI}$ calculated for $\delta=0$) has the following form
\begin{equation}
\begin{aligned}
&j_0\left(t\right)= 1-e^{-\frac{3\gamma t}{2}}
\left( \cosh \left(\digamma t\right)+ \frac{3\gamma}{2\digamma} 
\sinh \left(\digamma t\right) \right) ~, \\
&\oI_0 =\frac{\gamma\Omega ^2}{2\gamma ^2+\Omega ^2}~, \quad 
\digamma = \sqrt{\frac{\gamma^2}{4}-\Omega^2} ~,
\end{aligned}
\end{equation}
which matches expressions written for example in \cite{Carmichael1976,Kimble1977,Dagenais1978,Mandel1979}. Figure \eqref{fig:DagenaisFig4} shows this function drawn versus the dimensionless time $\gamma t$ for two values of $\Omega$. They match those drawn on Fig.4 of ref.\cite{Dagenais1978} calculated from equivalent formula and for the same parameters and also shown there to agree with the results of experiments on the resonance fluorescence of laser-excited sodium atoms in an atomic beam. Functions $j$ defined analogously for non resonant excitation are shown on Fig.\eqref{fig:DagenaisFig5} and they match those drawn on Fig.5 of ref.\cite{Dagenais1978}.

Differences have been noticed between the predictions in \cite{Pomeau2016,Pomeau2021,Pomeau2022} and those of the present method, with these differences large in the weak-excitation limit where the Rabi frequency $\Omega$ remains small compared to $\gamma\,, \delta$ (see Fig.2 in ref.\cite{Pomeau2022}). This is why we discuss also this limit, in which a good approximation is obtained by a perturbative expansion of the correlation function in terms of photon scattering amplitudes \cite{Dalibard1983}
\begin{equation}
\begin{aligned}
&j_\mathrm{pert}\equiv J_\mathrm{pert}/{\oI_\mathrm{pert}}
= \left|1-e^{-\left(\gamma-\imath\delta\right) t}\right|^2 ~, \\
&\oI_\mathrm{pert}
=\frac{\gamma\Omega ^2}{2\left(\gamma ^2+\delta ^2\right)}  ~.
\end{aligned}
\label{eq:photonscattering}
\end{equation}
The amplitudes interfering in eq.\eqref{eq:photonscattering} are two-photon scattering amplitudes which are sufficient to describe the fluorescence process in this limit. They correspond to two independent elastic Rayleigh scatterings (each linear in the field amplitude) on one hand, and one inelastic two-photon scattering (quadratic in the field amplitude) on the other hand. Again, this expression matches the results written above in the weak-excitation limit.

In this limit, the fluorescence flux is weak and experiments are more delicate than those corresponding to a higher flux. There are however configurations directly related to this discussion for which experiments have been performed and have confirmed the scattering formalism. When filters are used to select the inelastic scattering spectral lines, the interference leading to antibunching is destroyed but a bunching signal remains which is characteristic of the inelastic two-photon fluorescence process \cite{CohenTannoudji1979,Reynaud1981}. This signal has been observed experimentally and found to agree with the theory \cite{Aspect1980} (see also \cite{LopezCarreno2018,Phillips2020,Hanschke2020} for recent discussions of correlations between frequency filtered photons).
Another configuration corresponds to multiatom resonance fluorescence where photon antibunching has been observed in the field emitted coherently by many atoms under conditions of phase-matching \cite{Grangier1986} and found to agree with the theory \cite{Heidmann1985}.

\section{Conclusion}
\label{sec:discussion}

In this paper, we have reviewed the understanding of the radiative cascade of resonance fluorescence photons emitted by a laser-excited two-level atom.
We have considered the delay function as the primary characterization of this random point process, and derived other statistical characterizations from it. 

This method is perfectly adapted to the discussion of the \emph{quantum jumps} observed when fluorescence is interrupted by shelving the atom on a long-lifetime trap level \cite{Dehmelt1982}. The experimental observation of these quantum jumps  \cite{Nagourney1986,Sauter1986,Bergquist1986,Finn1986,Itano2015} was accompanied by a number of theoretical papers discussing this spectacular effect 
\cite{Cook1985,Javanainen1986,Pegg1986,Schenzle1986,Kimble1986,Zoller1987}.
The description in terms of momentary interruption of the radiative cascade of the dressed atom gave an extremely efficient and intuitive comprehension of intermittent fluorescence \cite{CohenTannoudji1986,Dalibard1987}.

Here, we considered only signals built on the fluorescence intensity, so that the process can be fully understood as a series of random times of emission of successive photons. The series can be drawn from independent random delays characterized by the function $K$ or, equivalently, by $P$, $\Lambda$ or $\lambda$.
The intrinsic randomness of the process is also emphasized in the Monte-Carlo wave-function approach to dissipative spontaneous emission which allows one to study other observables \cite{Dalibard1992,Dum1992,Molmer93,Plenio1998,Daley2014,Xu2015}.

\smallskip
\emph{Acknowledgements: }
This paper is dedicated to the memory of Jean Ginibre who initiated with Yves Pomeau and Martine Le Berre stimulating discussions about the intrinsic randomness of resonance fluorescence emitted by a single atom. I also want to thank Claude Cohen-Tannoudji and Jean Dalibard for many discussions spread over years of fruitful collaboration.

\end{document}